# Bigravity: a bimetric model of the Universe with variable constants, including VSL (variable speed of light)

**Jean-Pierre Petit and Gilles d'Agostini**
jppetit1937@yahoo.fr

______________________________________________________________________________

**Abstract**
The universe, far from being homogenous, expands in large empty bubbles of large-scale structure, but not in mass concentrations like galaxies, so that the Robertson-Walker solution does not fit. We suggest that a symmetry breaking occurred in a distant past, during the radiation-dominated era. Before, the three-dimensional hypersurface was invariant under the action of O(3) and the Robertson-Walker solution could be used. But this obliges the so called constants of physics, length and time scale factors, to be involved through a generalized gauge process, which is thus built. The subsequent variation of the speed of light solves the horizon problem, while the Planck barrier disappears.

______________________________________________________________________________

The present work is fairly different from other attempts published on the now so-called VSL, variable speed of light. The reader will find these other papers mentioned in the references at the end of -this paper.

1) **Introduction**

Let's present the general idea. The classical cosmological model was built from the Robertson-Walker solution of the Einstein field equations. This solution is based on the cosmological principle assuming the Universe is homogeneous and isotropic. Initially the model-makers believed that they could consider the Universe as a gas whose molecules would be the galaxies. In clusters these galaxies have a random velocity which may be compared to the thermal velocity of the kinetic theory of gases. The order of magnitude of the random galaxy velocities, with respect to the galaxy clusters, is around 1000 km/s, which is small if compared to the speed of light. So that theoreticians thought that this velocity could be neglected, this cosmic fluid being compared to dust ("dust Universe"). This was widely confirmed by observations.
Oppositely the observation of the large-scale structure gave evidence that matter was arranged around big voids, 100 light-years across in the mean, to be compared to joined soap bubbles". That was frankly non-homogeneous. As a consequence the curvature field is non-uniform.
A puzzling problem arises. Astronomers measure redshifts and conclude that the Universe is expanding, according to Hubble's law. But where does this expansion occur? Does the solar system expand? No. If it were expanding, it would be unstable. Do the galaxies expand? No, for the same reason.
To explain the measured redshifts we must admit some regions expand in the Universe and some others do not. Basically, the Robertson-Walker metric cannot take into account this non-homogeneity. In the Robertson-Walker metric we find a length scale factor *R* which depends only on the time-marker $x°$. It does not depend on space coordinates. It is supposed to be constant over the whole three-dimensional hypersurface $S(x°)$ at a given instant $x°$. It does not fit the observations so that we should think about a length scale factor which would depend on time and space. At microscopic scale we find cosmological, primeval photons forming the CMB, the cosmic microwave background radiation. Let us write their average wavelength as $\lambda$. It expands like the length scale factor $R$. The great voids are filled by such photons. Here is the expansion of the Universe. Photons behave like oscillations moving on an expanding cloth.

A material particle, whose mass is $m$, is associated ~~to~~ with a characteristic Compton~~'s~~ length:
(1)
$$\lambda_c = \frac{h}{mc}$$

If we consider the Planck constant $h$, the speed of light $c$ and the mass $m$ are invariant, this Compton~~'s~~ length does not vary in time. From this point of view, photons expand but matter doesn't. This corresponds to an idea introduced by Mach in 1883. In 1990, I illustrated this idea using a didactic image, in my book *"The Chronologicon"* from the series *"The Adventures of Archibald Higgins"*. This is the corresponding part, page 61:

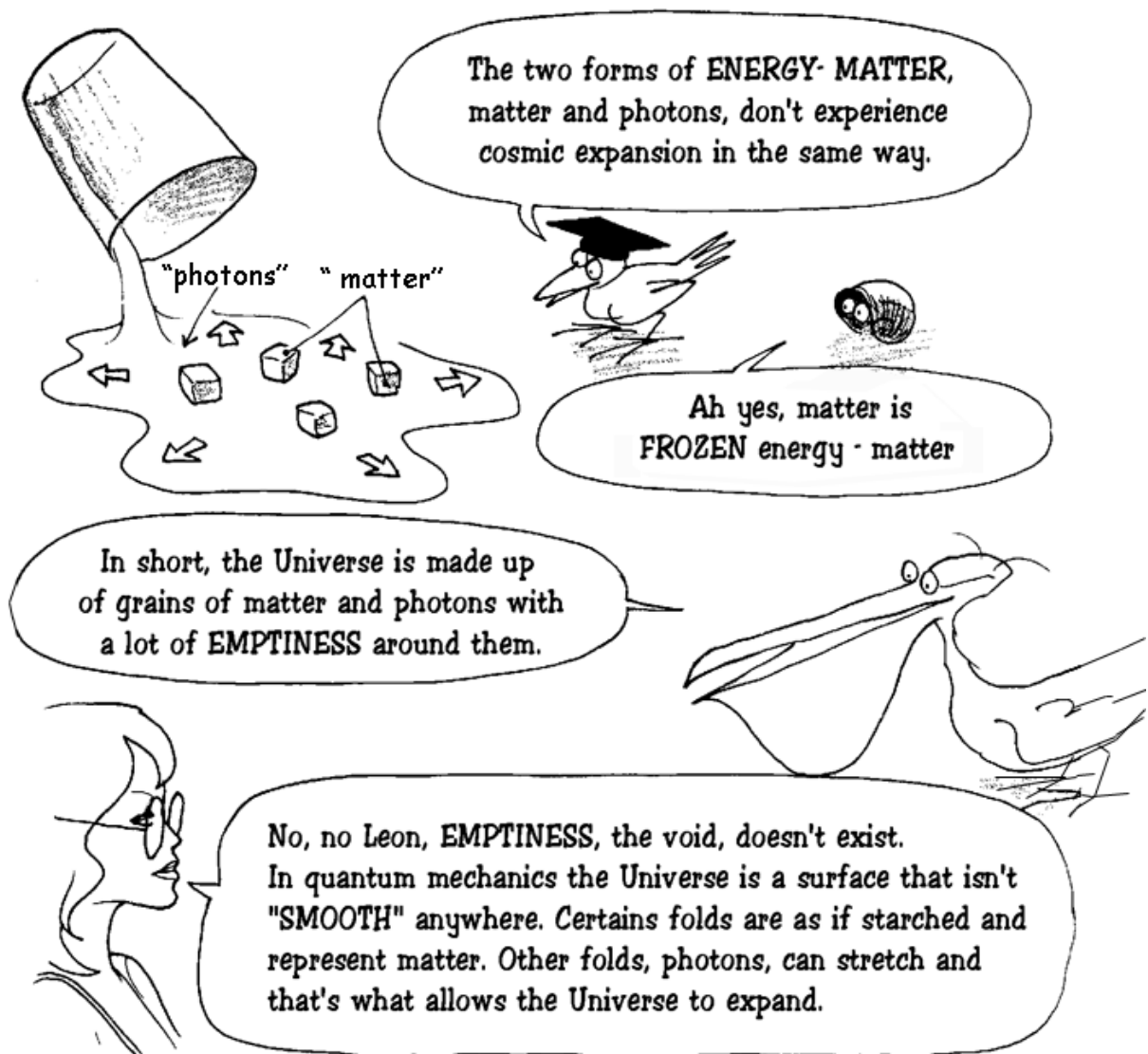

Figure 1: **Page 61 of my book "The Chronologicon"**

When we travel backwards in time, the Universe becomes hotter and hotter and, quoting Steven Weinberg in his famous book *"The first three minutes"*, it is "a mixture of all kinds of radiations". At this time, based on the CMB observation, the Universe looks very homogeneous. If we keep in mind the image of ice cubes immersed in water, they would melt and produce an homogeneous mixture.

This suggests a symmetry breaking, occurring in distant time during the radiation-dominated era. Have a look at the Robertson-Walker metric. First we define Gaussian coordinates, which implies that the three-dimensional surface is "oriented in space" and "oriented in time".

- One assumes that there exists a global time-coordinate, a global time-marker $x°$

- Space is assumed to be locally isotropic.

- Any two points in a three-dimensional space belonging to a fixed value of the time-coordinate are equivalent.

We choose an arbitrary point of the three-dimensional space to be the origin of spherical coordinates $(r, \theta, \varphi)$.
Introducing a length scale factor $R(x°)$ we can write:
(2)
$$r = R(x°) \, u$$

where $u$ is an adimensional variable.

Then the Robertson-Walker line element becomes:
(3)
$$ds^2 = dx^{o\,2} - (R_{(x^o)})^2 \frac{du^2 + u^2(d\theta^2 + \sin^2\theta \, d\varphi^2)}{(1 + k^2 u^2)^2}$$

$k = \{-1, 0, +1\}$ is the curvature index. The coordinates $(u, \theta, \varphi)$ are pure numbers or angles. The hypothesis of isotropy and homogeneity gives a $R$-field which only depends on time. A point whose coordinates $(u, \theta, \varphi)$ are invariant is called co-moving (with space). As pointed out earlier this description does not fit ~~with~~ the observational data.

We can offer a 2d didactic image of such geometry. See figure 2. On the right we have drawn a cube with blunt vertices. Each vertex is one eighth of a sphere. The eight blunt corners are connected through portions (quarters) of cylinders and square flat surfaces, i.e. Euclidean elements. The two images 3 and 4 in figure 2 suggest the expansion of a closed Universe containing eight "mass concentrations" (the eight blunt vertices). The flat surfaces expand, not the curved corners. These keep a constant area. This is a didactic image of a closed Universe with mass concentration areas (that do not expand) separated by voids (the Euclidean portion of cylinders and flat squares).

The time-sequence is supposed to go from image 1 to image 4. In image 2 the eight portions of the sphere join together. Then this closed universe becomes a sphere, that has O(2) symmetry. We find a symmetry breaking from step 2 to step 3. Before, the O(2) symmetry holds. After, it doesn't.

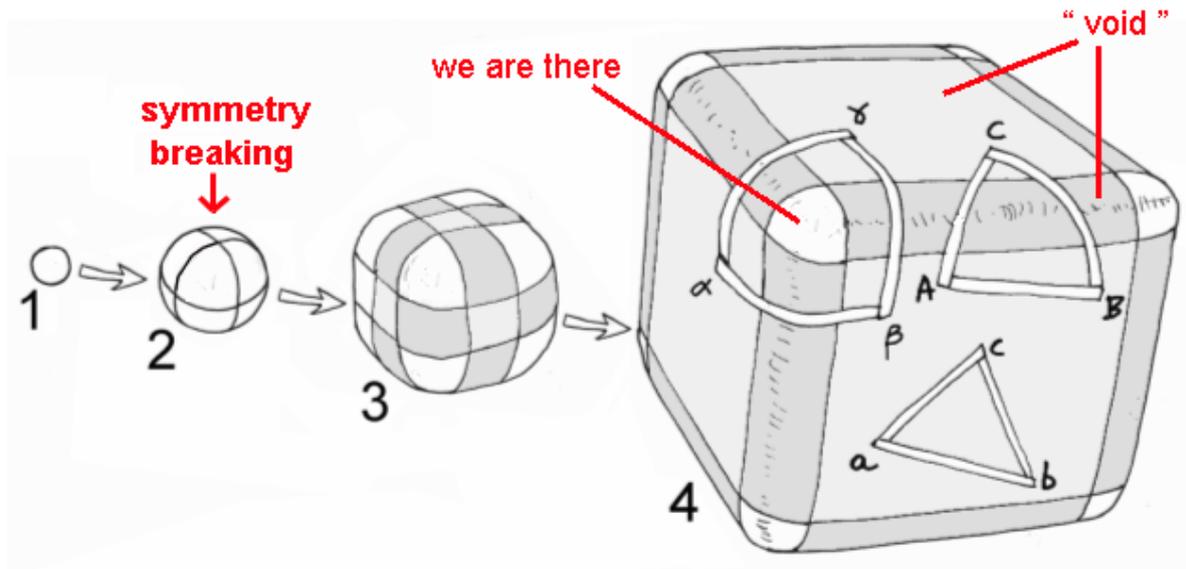

Figure 2: **Two-dimensional didactic image of an expanding, closed Universe experiencing a symmetry breaking in step 2.**

Geometers can immediately extend this for a three-dimensional closed surface, illustrated as a set of flat volumes joined to constant curvature portions of space. In the two-dimensional model we have shown eight "mass concentrations", but we could put as many as we want, thus forming a diamond-like object, with blunt vertices (rounded summits). Similarly, a 3-sphere could be transformed into a many-3d-faces 3d-diamond whose curved volumes would be smoothed (constant curvature and constant volume object) linked by euclidean elements.

Similarly we could imagine a symmetry breaking. Before the objects is a 3-sphere, having an O(3) symmetry. After, this symmetry is broken or lost. The object becomes a three-dimensional polyhedron where each constant curvature (and volume) portion represents a galaxy or some mass concentration that does not expand. .

This is the general idea. We suppose the Robertson-Walker metric is suitable to describe the (homogeneous) *early* Universe. Then a symmetry breaking occurs, very early in the radiation-dominated era. Matter concentrations begin to form. We suggest that this corresponds to two different evolution processes.

When we want to build a model of cosmic evolution we have to deal with:

- Pure geometric features, governed by Einstein's field equations

- Features linked to special relativity (invariance through Lorentz rotations)

- Electromagnetic features (the particles are linked by electromagnetic forces, ruled by Maxwell's equations)

- Quantification is present, which is governed by, for example, the (non relativistic) Schrödinger equation.

All these features come from local observations and experiments, which satisfy a certain set of physical equations, containing the following quantities:

> $G$: gravitational constant
> $c$: speed of light
> $m$: masses
> $h$: Planck constant
> $e$: elementary charge
> $\mu_o$: magnetic constant (vacuum permeability)

From this set of equations we can build measuring instruments. They extend our senses. But finally the last measuring instrument is ourselves, with our body as a length scale and our average life, a "shelf time", as a time scale. We may compare this duration to astronomical phenomena, like the year, the day (a complete rotation of the Earth). We can build mechanical systems, like a torsion pendulum, and discover that they can be used as clocks. Comparing this object to our body, our hand, we can build rules. We can divide it, and so on.

We replace the fully human measurement by mechanical measurements, based on the equations of physics. It works. We can perform measurements of constants and we don't find any change, so we think they should be absolute constants. Quantum mechanics works too and brings a new insight into the nature of matter. From the constant of today's physics we can build two characteristic quantities

The Planck length:
(4)
$$L_P = \sqrt{\frac{hG}{c^3}}$$

The Planck time:
(5)
$$t_P = \sqrt{\frac{hG}{c^5}}$$

Quantum mechanics generates its own limits in space and time. It becomes impossible, through QM formalism, to analyse or even to conceive processes occurring on lengths and durations shorter than these characteristic values.

As the scientist believes the constants found are absolute constants, that do not vary alongside cosmic evolution, he thinks of some hypothetical "quantum era" and asks how things could have been in the extremely early conditions.

2) **Existence of a fundamental gauge relation**

All physicists know the power of dimensional analysis. Considering a given set of physical equations they can introduce characteristic length and time and introduce adimensional space and time variables. Then they can weight, measure the relative importance of different terms in an equation in which the constants of physics appear.

This could be extended, considering that we can use today's values of the constants, considering they may vary, and introduce their adimensional form. As an example, let us consider the constant of gravitation. Call $G_o$ today's measured value. If we admit that G may vary, we can write:
(6)
$$G = \Gamma \, G_o$$

$\Gamma$ being an adimensional quantity. We can do similar operations for all constants, as we do for length and time. Now, ask the question:

- Can we find what we will call a *generalized gauge transform*, linking the length and time scale factors, and the adimensional variables, which takes into account the variation of the physical constants, and which would keep all the equations invariant?

The answer is yes. There is a single one, see reference [4]. Note that we cannot obtain evidence of such variations using our measuring instruments. The reason is very simple: these systems are precisely built from these quantities, so it becomes impossible to find evidence for any constant variation among all these constants of physics.

Similarly, if you wish to measure the dilatation of a table made of iron, using a ruler made of the same metal, due to ambient thermal variation you could not as the table and the scale would experience the same parallel variations.

We will present further this generalized gauge transform. If $R$ and $T$ are respectively the length and time scale factors, all the characteristic lengths of physics, including the Planck length, are found to vary like $R$. All the characteristic times of physics are found to vary like $T$. In addition, all the energies are conserved. This remarkable gauge property is based on a group, to be discovered.

3) **Today's physics**

No astronomer would pretend that the solar system or the galaxies expand with the Universe. Consider some sort of reference system composed by two masses *m* circling aroung their common center of gravity:

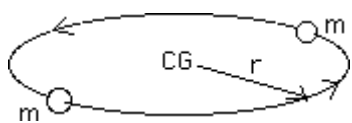

Figure 3: **Two masses orbiting around their common center of gravity**

The centrifugal force is counterbalanced by the gravitational attraction:

(7)
$$\frac{mV^2}{r} = \frac{Gm}{4r^2}$$

If the radius of this circular orbit was extended, the system would become unstable. In the classical vision of celestial mechanics, *G* and *m* are considered to be absolute constants. The kinetic energy and angular momentum are conserved, so that *V* does not change neither. This circular orbit could not be co-moving in a Robertson-Walker geometry, where the expansion occurs everywhere.

In general relativity, masses follows the geodesics of a four-dimensional space-time hypersurface. These masses cannot follow geodesics associated with the Robertson-Walker metric. We do not know, presently, how to build a solution of to the field equations which takes into account these features. When we make local calculations, for example for perihelion precession of Mercury or gravitational lensing, we use time-independent Schwarzschild solutions to the field equations.

Let us return to our didactic image: the cube with blunt vertices and smoothes edges. Our circular orbit takes place in the non-expanding zone, in a corner which does not expand.

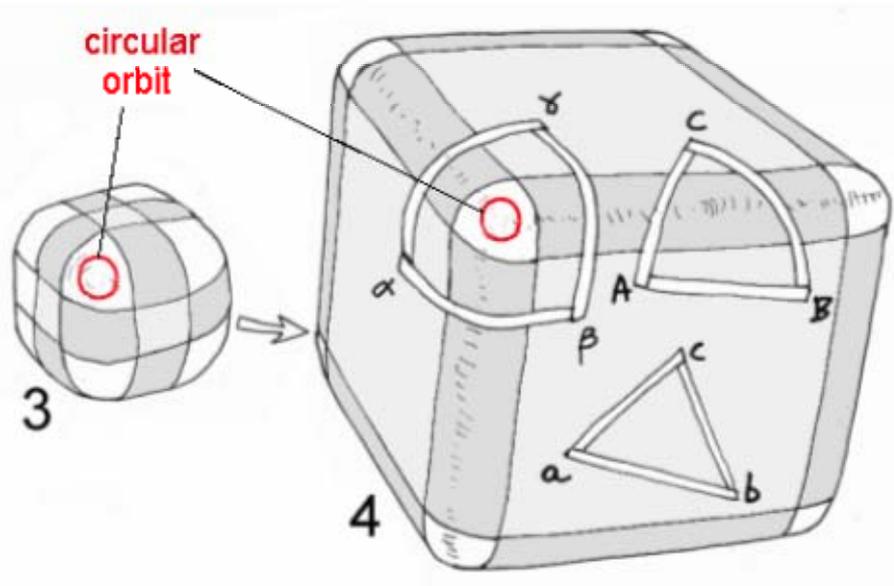

Figure 4: **Reference circular orbit location**

This is a composite geometry since curved and flat elements form the object. To be closer to reality, this circular orbit should be inscribed in a non-expanding volume, part of a three-dimensional hypersurface. There, it would be part of a region of space-time that would behave like a quasi steady-state element of the geometric solution, while the area between the mass concentrations space-time would be close to a Friedman non-steady solution. Currently we do not know how to manage that.

4) **The early evolution as a generalized gauge process**

If we go further backwards into the past, we meet the symmetry breaking event. Then the Robertson-Walker solution holds and the Universe obeys an O(3) symmetry. If we want to inscribe the circular orbit corresponding to our two masses linked by gravitational force, the span of this co-moving orbit must vary like $R(x^\circ)$. See the small red circle:

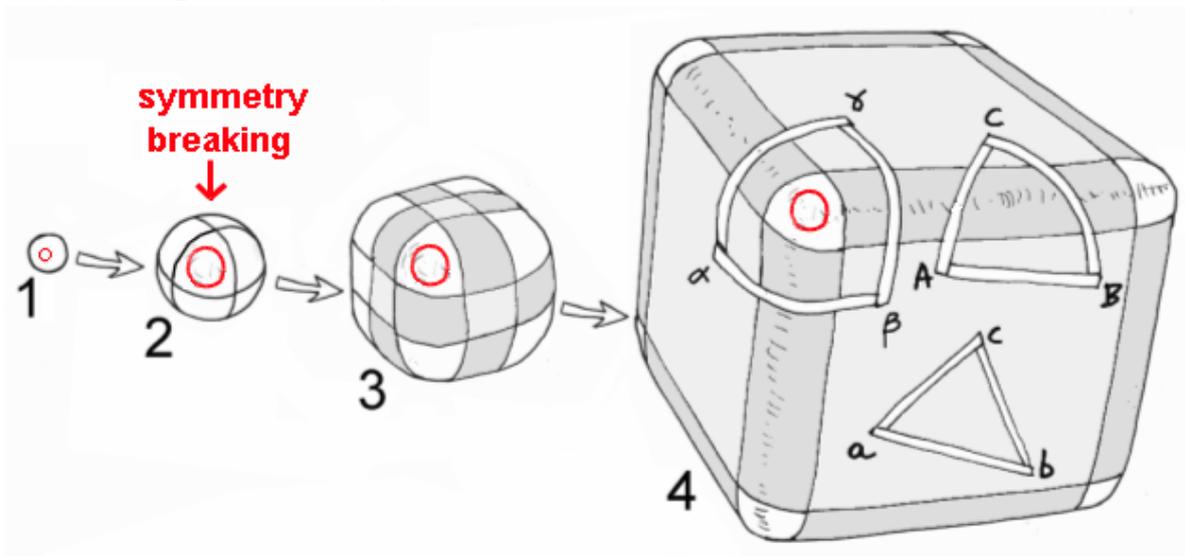

.
Figure 5: **2d didactic image of the symmetry breaking**

We will assume the Universe undergoes a generalized gauge process, keeping the equations invariant. At the end of this study the benefit will be the justification of the homogeneity of the early Universe with no need to call for the inflation theory. Let us build the gauge laws. As shown in [1] this gauge process make all the characteristic lengths of physics to follow the variation of the length scale factor $R$. The energies are conserved. Look at the invariance of the field equation, in which we take a cosmological constant equal to zero, while it does not refer to early Universe and short range gravitational interaction.

(8)
$$S = \chi\, T$$

It is divergenceless, which implies that the constant $\chi$ must be an absolute constant. This last is classically determined through an expansion into a series of the metric, the zero-order term being the time-independent Lorentz metric. We add a perturbation term which is also time-independent:

(9)
$$g = \eta + \varepsilon\, \gamma$$

We perform a Newtonian approximation (weak field, low velocities with respect to the speed of light). The constant is determined, identifying the linearized field equation to Poisson's equation. The expression depends on how we decide to write the tensor $T$. Let us follow [20], section 10.5:

(10)
$$T^{oo} = \rho$$

Then the identification gives:

(11)
$$\chi = -\frac{8\pi G}{c^2}$$

The Einstein equation is invariant if
(12) $$G \approx c^2$$
Notice that, when we determine the expression of Einstein's constant combining $G$ an $c$ in (11), this does not imply that these two should be absolute constants, for the perturbation method is based on time-independent terms in the expansion into a series of the metric.
Writing that the Schwarzschild length varies like $R$, we get:

(13) $$\frac{Gm}{c^2} \approx R \qquad m \approx R$$

The conservation of the energy brings immediately:
(14) $$mc^2 \approx Cst \qquad c \approx \frac{1}{\sqrt{R}} \qquad G \approx \frac{1}{R}$$

If we write the Planck length varies like $R$ we get:
(15) $$\sqrt{\frac{hG}{c^3}} \approx R \qquad \frac{h}{c} \approx R^2 \qquad h \approx R^{3/2}$$

The invariance of the kinetic energy (or the fact that the Jeans length varies like $R$) gives:
(16) $$V \approx \frac{1}{\sqrt{R}} \approx c$$

If we write that the radius of the circular orbit of the two masses orbiting around their common center of gravity varies like $R$, we get:
(17) $$T \approx R^{3/2}$$

Remark that:
(18) $$R = cT$$

Writing that the energy $h\nu = h/\tau$ is conserved we obtain:
(19) $$\tau \approx h \approx R^{3/2} \approx T$$

Notice this looks like a first link between quantum and gravitational worlds. Let us look now to electromagnetism. Write the Bohr radius varying like R:
(20) $$r_b = \frac{h^2}{m_e e^2} \approx R \qquad e \approx \sqrt{R}$$

We can complete that, assuming that the fine-structure constant $\alpha$ does not vary in this gauge evolution process:
(21) $$\alpha = \frac{e^2}{\varepsilon_o h c} \qquad \varepsilon_o = \text{Constant} \qquad \mu_o \approx R$$

This would be a consequence of the invariance of Maxwell's equations, coupled to the hypothesis of ~~the~~ electromagnetic energy conservation [26]. Of course we would find that all the characteristic lengths of electromagnetism (like the Debye length) vary like *R,* while the coulomb cross section varies like $R^2$. Similarly all the characteristic times of physics are found to vary like the time scale factor *T*. This gives one gauge variation law keeping all equations invariant.

As we said, our goal is to justify the observed homogeneity of the early Universe without calling on inflation theory to rescue the model. At this level we must start from the results of our bimetric model, presented in an earlier paper [18]. We recall that the motivation then was to clarify the nature of so-called "dark energy", producing the late acceleration of the expansion process. This "component" of the Universe was identified with a second kind of mass and photons, possessing negative energy (notice that this bimetric model has nothing to do with other authors' works using this technique).

5) **Link with our bimetric model**

Let us give the basis of this bimetric model. We assume that the Universe contains two kinds of particles. The first, with positive or zero mass and positive energy, follow the geodesics corresponding to a first metric $g^+$. The second ones, with negative or zero mass and negative energy, follow the geodesics built from a second metric $g^-$. As explained in the reference mentioned, the two metrics are coupled through the following field equations:

(22) $$S^+(g^+) = \chi ( T^+ - T^- )$$

(23) $$S^-(g^-) = \chi ( T^- - T^+ )$$

Assuming the Universe to be isotropic and homogeneous, at very large scale we introduce Robertson-Walker metrics, with their own length and time scale factors $R^+$ and $R^-$, which gives the two coupled differential equations:

(24)
$$\frac{d^2 R^+}{dx^{\circ 2}} = -\frac{1}{(R^+)^2}\left(1 - \frac{(R^+)^3}{(R^-)^3}\right)$$

$$\frac{d^2 R^-}{dx^{\circ 2}} = -\frac{1}{(R^-)^2}\left(1 - \frac{(R^-)^3}{(R^+)^3}\right)$$

The obvious divergence of the solution has been evoked to construe the observed acceleration. As the time-marker $x^\circ$ grows, the solutions become more and more divergent. The negative energy component behaves like repellent "dark energy" and accelerates our positive energy matter. But when going backwards in time, the two scale factors get the same value and follow a linear evolution in $x^\circ$. If it is identified with a cosmic time *t*, through $x^\circ = ct$, *c* being considered as an absolute constant, this evolution becomes so slow that all the hydrogen of the Universe would be converted into helium.

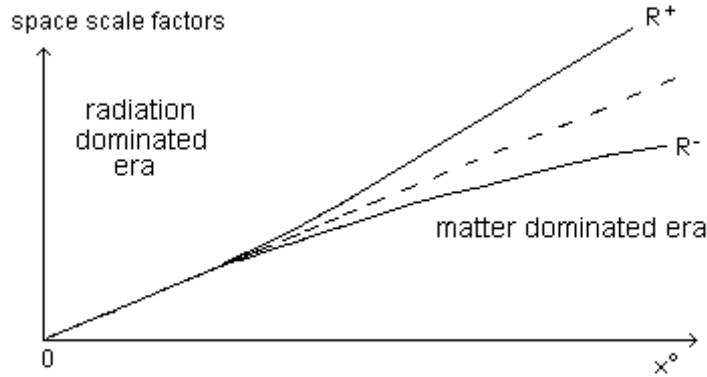

Figure 6: **The bimetric cosmological model ( $R^+$ , $R^-$ ) as functions of the time-marker $x°$**

In figure 7 we have represented the evolution of our reference system, of two masses coupled by gravitation, with the time-marker $x°$. At the very begining the radius of the orbit follows the growth of the length scale factor *R*. The two geodesics spiral along and around a cone. Then the symmetry breaking occurs. The portion of space where the circular orbit is located is in a non-expanding region: the geodesics spiral along the length of a cylinder. We have illustrated in red the world-lines of co-moving particles.

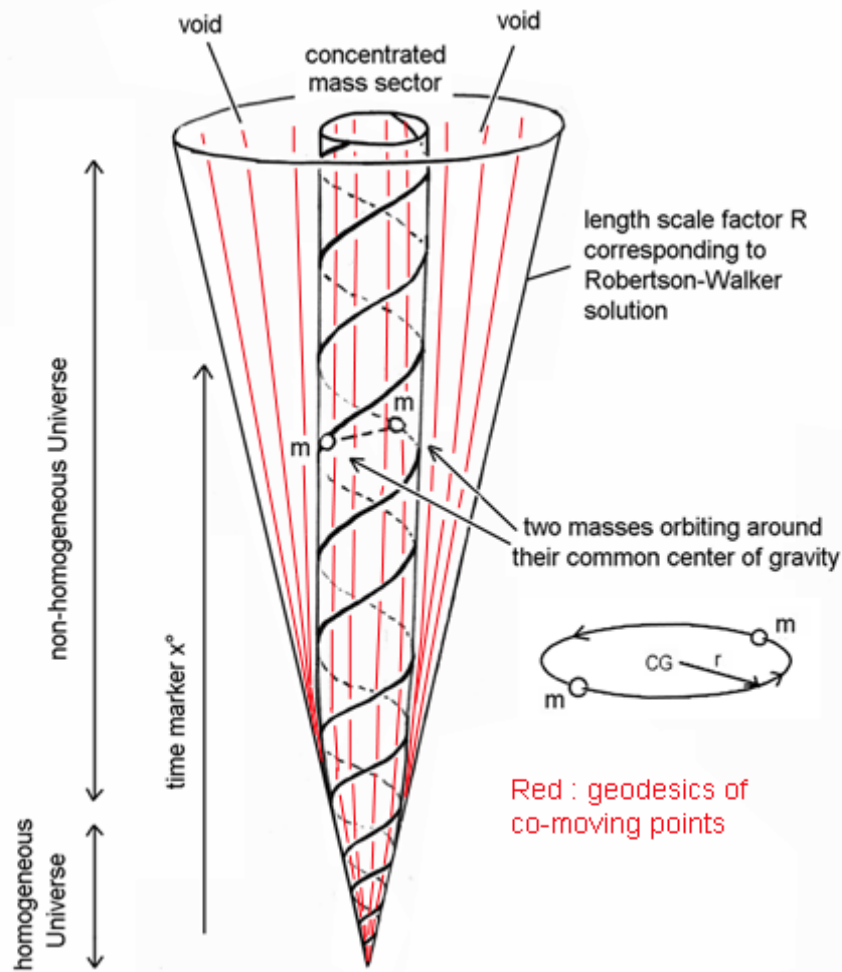

Figure 7: **Red: world-lines (geodesics) of co-moving particles. Black: geodesics paths of the two masses linked by gravitational force. Before the symmetry breaking the Universe is homogeneous and the orbit grows like *R*. Then its radius becomes constant.**

Before the symmetry breaking the (common) metric is that of Robertson-Walker. We take k = 0 which takes flatness into account.
(25)
$$ds^2 = dx^{o\,2} - [R_{(X^o)}]^2 \left[ du^2 + u^2 (d\theta^2 + \sin^2\theta\, d\varphi^2) \right]$$

Starting from a linear evolution of $R$ versus $x^o$ we put the metric into a conformally flat form:
(25)
$$ds^2 = [R(\tau)]^2 \left[ d\tau^2 - du^2 - u^2 (d\theta^2 + \sin^2\theta\, d\varphi^2) \right]$$

introducing the new time-marker $\tau = Log\, x^o$. Now we introduce the time scale factor $T(\tau)$
(26)
$$ds^2 = [c(\tau)]^2 [T(\tau)]^2 d\tau^2 - [R(\tau)]^2 \left[ du^2 + u^2 (d\theta^2 + \sin^2\theta\, d\varphi^2) \right]$$

Notice the null geodesics are basically gauge-invariant.
Consider a given value of the time-marker $\tau_o$ and an interval $\Delta\tau$ small enough to make possible to consider the functions $c(\tau)$, $T(\tau)$ et $R(\tau)$ as invariant. The line element can be written:
(26):
$$ds^2 = [C(\tau_o)]^2 [T(\tau_o)]^2 d\tau^2 - [R(\tau_o)]^2 \left[ du^2 + u^2 (d\theta^2 + \sin^2\theta\, d\varphi^2) \right]$$

the functions being linked through:
(27)
$$R(\tau_o) = C(\tau_o)\, T(\tau_o)$$

During this short interval $(\tau_o, \tau_o + \Delta\tau)$ write:
(28)
$$t = T(\tau_o)\, \tau$$
(29)
$$r = R(\tau_o)\, u$$

We choose (28) as a definition of the physical time. During this small time interval $(\tau_o, \tau_o + \Delta\tau)$ we can write the following line element:
(30)
$$ds^2 = [c(\tau_o)]^2 dt^2 - dr^2 - r^2 (d\theta^2 + \sin^2\theta\, d\varphi^2)$$

Using Cartesian coordinates:
(31)
$$ds^2 = [c(\tau_o)]^2 dt^2 - (dx^1)^2 - (dx^2)^2 - (dx^3)^2$$

Considering the null geodesics, we obtain the value $c(\tau_o)$ of the speed of light at the instant $\tau_o$

6) **Solving the problem of the cosmological horizon**

Imagine that light propagates along the $x$ direction. We have:
(32)

$$dx = c(\tau) dt$$

Express all as functions of the length scale factor *R*:
(33)
$$c \approx \frac{1}{\sqrt{R}} \qquad dt = T(\tau) d\tau \qquad d\tau = \frac{dx^\circ}{x^\circ} = \frac{dR}{R} \qquad T \approx R^{3/2}$$

The horizon is given by the integral:
(34)
$$horizon = \int c\, dt = \int_0^R C(\xi)\, T(\xi) \frac{d\xi}{\xi} = \int_0^R \frac{1}{\sqrt{\xi}} \xi^{3/2} \frac{d\xi}{\xi} = \int_0^R d\xi = R$$

It varies like the length scale factor *R*, which ensures the homogeneity of the Universe at any time.
As all this model goes with a Planck length varying like *R* (while the Planck time varies like *T*) we see that the Planck barrier disappears. The creation of this Planck barrier is due to the hypothesis that the constants of physics do not vary. Then we project the local and present aspect of microphysics towards the most distant past.

7) **About time**

What is *time* when we consider the very early state of the Universe, when all the particles cruise at relativistic velocities? How can we build a physical clock? If that is not possible, what is the meaning of a time that no one could measure?

Let us go back to our system composed of two masses circling around their common center of gravity, now considered as an elementary clock. There is no absolute measurement of time, only a relative measure, a comparison of a duration with respect to a reference one. In any case a turn represents some sort of time measurement. Following the breadcrumbs trail that is the chronological time-marker $x^\circ$ we can count how many turns occur from a given instant $X^\circ$ to the value zero of $x^\circ$.
The period is t:
(35)
$$period = \frac{2\pi r^{3/2}}{Gm}$$

But:
(36)
$$Gm = constant \qquad r \approx R \qquad period \approx R^{3/2} \approx x^{\circ\, 3/2}$$

During a certain interval $dx^\circ$ of the chronological time-marker the number of turns increases by:
(37)
$$dn = \frac{dx^\circ}{x^{\circ\, 3/2}}$$

Between $x^\circ = 0$ and $x^\circ = X^\circ$ the number of turns is:
(38)

$$n = \int_0^{x^°} \frac{dx^°}{x^{°3/2}} = \left[\frac{1}{\sqrt{x^°}}\right]_0^{x^°} = \text{infinite}$$

We may consider a turn of this "elementary clock" as an "elementary event". From a mathematical point of view, the Robertson-Walker solution starts from a zero value of the length scale factor *R*. The count of the number of turns of our "elementary clock" evokes well known Zeno's paradoxes.

What could be the meaning of such a result? It seems to mean that when we go backwards in time, towards what we consider as an origin or singularity, an infinite number or "elementary events" occurs. Then the universe looks like a very peculiar book. The chronological time-marker $x^°$ is the width of the book. As we flick through it to get back to the beginning, its pages get thinner and thinner. An infinity of pages needs to be flicked through to get to the start of the beginning, and we can never read the author's preface.

8) **The evolution of the constants of physics**

For the early Universe we have assumed that a symmetry breaking occurred during the radiation-dominated era. Then the evolution was phrased through equations (12) to (21). A question arises. If there is a phase transition, when does it occur and why? We cannot answer the question at the present time. If we express the generalized gauge process, choosing the density of electromagnetic energy (which is dominant) as the "leading parameter", we obtain:
(39)

$$G \approx \sqrt{\rho} \quad m \approx m_e \approx \frac{1}{\sqrt{\rho}} \quad h \approx \rho^{-3/4} \quad c \approx v \approx \rho^{1/4} \quad \mu_o \approx \frac{1}{\sqrt{\rho}}$$

$$R \approx \frac{1}{\sqrt{\rho}} \quad T \approx \rho^{-3/4} \quad E \approx \rho^{-3/4} \quad B \approx \sqrt{\rho} \quad e \approx \rho^{-1/4}$$

Just to fix ideas we can introduce some function which involves/some functions which involve a critical value of this density: (40)

$$G = G_o \sqrt{\vartheta(\rho)} \quad m = m_o \frac{1}{\sqrt{\vartheta(\rho)}} \quad h = h_o \left[\vartheta(\rho)\right]^{-3/4}$$

$$c = c_o \left[\vartheta(\rho)\right]^{1/4} \quad e = e_o \left[\vartheta(\rho)\right]^{-1/4}$$

with:
(41)
$$\vartheta(\rho) = 1 + \frac{\rho}{\rho_{cr}} = 1 + x$$

we get:

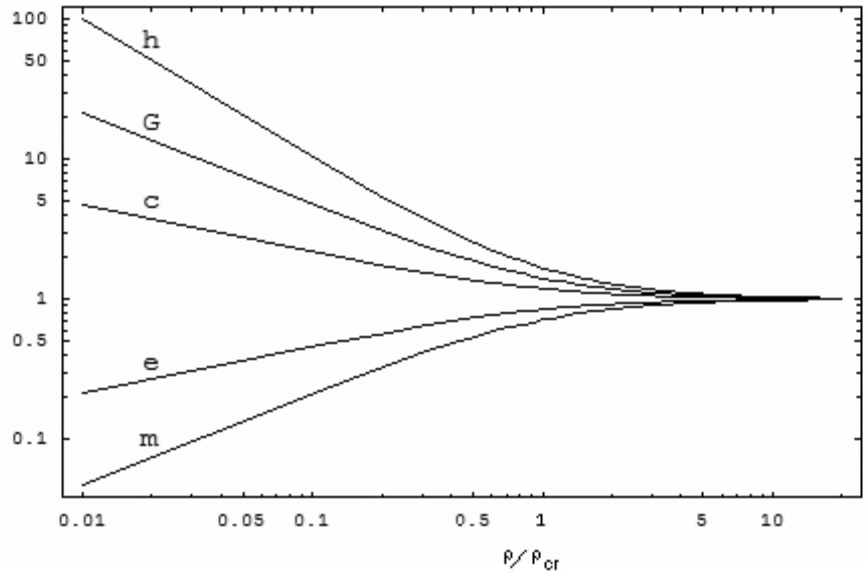

Figure 3: **Compared evolutions of the constants of physics in the early universe**

There are regions in the Universe where density can reach extremely high values: in the cores of neutron stars (in such a place it is as if the universe's evolution was going backward in time). The state of the matter in the core of the star is similar to that of the very early Universe). It could be interesting to consider the evolution of this star core beyond a certain density value, as accompanying an alteration of local physical constants, as local length and time scale factors. We speculate that this could modify the topology inside the core, bearing a hypertoric space bridge suitable for transfering excess matter to another Universe. Another paper will discuss this possibility, giving insights into the first work undertaken on this idea.

**Conclusion**

We started from the remark that the Robertson-Walker solution, assuming the cosmos is isotropic and homogeneous, does not describe the observed universe since it is clearly non-homogeneous, and that expansion is supposed not to occur within the vast regions occupied by galaxies. We suggested this non-homogeneity was born from a symmetry breaking that happened in a distant past, where the three dimensional hypersurface lost its symmetry O(3).

We proposed the hypothesis that masses, and in particular a couple of masses *m* linked by gravitational force and orbiting around their common center of gravity, follow the geodesics of the four-dimensional space-time hypersurface. We have shown that this is possible, if the equations of physics are to keep their validity, only if the physical constants undergo joint variations.

We established the gauge variation laws linking them, as well as the length and time scale factors. We obtained a description of cosmic evolution where the cosmological horizon varies like the length scale factor, which guarantees the homogeneity of the early Universe without calling on the inflation theory.

In this generalized gauge process, before the symmetry breaking occurs, all characteristic lengths of physics vary like the length scale factor *R*, while all characteristic times vary like the time scale factor *T*. In this way the Plank barrier disappears.

Focussing on our two mass system linked by gravitational force, we assimilated it to an elementary clock where time would be measured by each turn count. After the symmetry breaking, the number of turns evolves like classical cosmic time, following from $x° = ct$ with $c$ being an absolute constant. We have shown that in the previous era, before this symmetry breaking, the elementary clock makes an infinite number of turns which brings up the problem of the definitions of "time" and of the "origin".

**Acknowledgments :** The authors thank Dr B. Kolev for useful advices and comments, J.Geffray and John Murphy for the translation of the paper into English.

# References

**[1]** J.P. Petit (Nov. 1988). "An interpretation of cosmological model with variable light velocity". Modern Physics Letters A, 3 (16): 1527.
**[2]** J.P. Petit (Dec. 1988). "Cosmological model with variable light velocity: the interpretation of red shifts". Modern Physics Letters A, 3 (18): 1733.
**[3]** J.P. Petit, M. Viton (1989). "Gauge cosmological model with variable light velocity. Comparizon with QSO observational data". Modern Physics Letters A, 4 (23): 2201–2210.
**[4]** J.P. Petit (1995). "Twin Universes Cosmology". Astronomy and Space Science (226): 273–307.
**[5]** A. Albrecht, J. Magueijo (1999). "A time varying speed of light as a solution to cosmological puzzles". Phys. Rev. D59. arXiv:astro-ph/9811018
**[6]** J. Magueijo (2003). "Faster Than the Speed of Light: The Story of a Scientific Speculation". Perseus Books Group , Massachusetts. ISBN 0738205257
**[7]** J.W. Moffat (1993). "Superluminary Universe: A Possible Solution to the Initial Value Problem in Cosmology". Int. J. Mod. Phys. D2, 351–366. arXiv:gr-qc/9211020
**[8]** J.D. Barrow (July 24, 1999). "Is nothing sacred?". New Scientist 163 (2196): 28–32.
**[9]** C.B. Netterfield *et al.* (2002). "A measurement by BOOMERANG of multiple peaks in the angular power spectrum of the cosmic microwave background". Astrophys. J. (571): 604-614. arXiv:astro-ph/0104460
**[10]** C.L. Bennet *et al.* (2003). "First Year Wilkinson Microwave Anisotropy Probe (WMAP) Observations: Preliminary Maps and Basic Results". Astrophys.J.Suppl. 148. arXiv:astro-ph/0302207
**[11]** G. Amelino-Camelia (2002). "Doubly-Special Relativity: First Results and Key Open Problems". Int. J. of Mod. Phys. D11, 1643. arXiv:gr-qc/0210063
**[12]** J.D. Barrow & J. Magueijo (2000). "Can a Changing α Explain the Supernovae Results?". Astrophys. J. Lett. 532, L 87–90.
**[13]** M.A. Clayton, J.W.Moffat (1999). Phys. Lett. B (460): 263-270. arXiv:astro-ph/9907354
**[14]** S. Alexander (2000). "On The Varying Speed of Light in a Brane-Induced FRW Universe". JHEP 0011 017, arXiv:hep-th/9912037
**[15]** J. Magueijo (2001). "Stars and black holes in varying speed of light theories". Phys. Rev. D63, 043502.
**[16]** J. Magueijo, L. Smolin (Dec. 18, 2001). "Lorentz invariance with an invariant energy scale". Phys.Rev.Lett. (88) 190403. arXiv:hep-th/0112090
**[17]** B.A. Basset, S. Liberati, C. Molina-Paris, M. Visser (2000). "Geometrodynamics of Variable Speed of light Cosmologies". Phys.Rev. D62 103518. arXiv:astro-ph/0001441
**[18]** J.P. Petit, G. d'Agostini (August 2007). "Bigravity as an interpretation of the cosmic acceleration". CITV (Colloque International sur les Techniques Variationnelles, tr. *International Meeting on Variational Techniques*), Le Mont Dore. arXiv:math-ph/0712.0067
**[19]** J.P. Petit, G. d'Agostini (August 2007). "Bigravity: A bimetric model of the Universe. Positive and negative gravitational lensings". CITV (Colloque International sur les Techniques Variationnelles, tr. *International Meeting on Variational Techniques*), Le Mont Dore. arXiv:math-ph/0801.1477
**[20]** R. Adler, M. Bazin, M. Schiffer (1975). "Introduction to General Relativity", Mc-Graw Hill Books Cie, 2nd edition. ISBN 0070004234
**[21]** A.H. Guth (1981). "Inflationary universe: A possible solution to the horizon and flatness problems". Phys. Rev. D 23 (2):347–356.
**[22]** A.D. Linde (1982). "A new inflationary universe scenario: A possible solution of the horizon, flatness, homogeneity, isotropy and primordial monopole problems". Phys. Lett. B108, 1220.
**[23]** A. Albreicht, P. Steinhardt (April 1982). "Cosmology for Grand Unified Theories with Radiatively Induced Symmetry Breaking". Phys. Rev. Lett. 48 (17):1220–1223.
**[24]** A.D. Linde (1983). "The inflationary Universe". Phys. Lett. B 129, 177179.
**[25]** J.P. Petit (July 1994). "The missing mass problem". Il Nuovo Cimento B, 109: 697–710.
**[26]** J.P. Petit, P. Midy, F. Landsheat (June 2001). "Twin matter against dark matter". Intern. Meet. on Astrophys. and Cosm. *"Where is the matter?"*, Marseille, France.